\documentclass[preprintnumbers,showpacs,amsmath,amssymb,floatfix,prd,twocolumn,superscriptaddress,nofootinbib]{revtex4}
\usepackage{bbm}
\usepackage{graphicx}
\usepackage{epsfig}
\usepackage{bm}
\usepackage{amsfonts}
\usepackage{epstopdf}
\usepackage{multirow}

\begin{document}

\title{Global behavior of cosmological dynamics with interacting Veneziano ghost}

\author{Chao-Jun Feng}
\email{fengcj@shnu.edu.cn} \affiliation{Shanghai United Center for Astrophysics (SUCA), \\ Shanghai Normal University,
   100 Guilin Road, Shanghai 200234, P.R.China}

\author{Xin-Zhou Li}
\email{kychz@shnu.edu.cn} \affiliation{Shanghai United Center for Astrophysics (SUCA),  \\ Shanghai Normal University,
   100 Guilin Road, Shanghai 200234, P.R.China}

\author{Ping Xi}
\email{xiping@shnu.edu.cn} \affiliation{Shanghai United Center for Astrophysics (SUCA),  \\ Shanghai Normal University,
    100 Guilin Road, Shanghai 200234, P.R.China}

\begin{abstract}
In this paper, we shall study the dynamical behavior of the universe accelerated by the so called Veneziano ghost dark energy  component locally and globally by using the linearization and nullcline method developed in this paper. The energy density is generalized to be proportional to the Hawking temperature defined on the trapping horizon instead of Hubble horizon of the Friedmann-Robertson-Walker (FRW) universe. We also give a prediction of the fate of the universe and present the bifurcation phenomenon of the dynamical system of the universe.  It seems that the universe could be dominated by dark energy at present in some region of the parameter space.
\end{abstract}


\maketitle

\section{Introduction}\label{sec:intro}
Currently, lots of theoretical  models were proposed to explain the accelerating of the universe, which is convinced by many observations. In some of the models, people proposed a new kind of dark energy component with negative pressure  in our universe, which will drive the acceleration, and the simplest dark energy model is the cosmological constant, but it suffers fine-tuning and coincidence problems. While, in other models, people try to modify the Einstein gravity at large scale in the universe, e.g. $f(R)$, DGP, etc. model, then the universe can be accelerated without introducing dark energy.

Recently, a very interesting dark energy model called Veneziano ghost dark energy has been proposed \cite{Urban}, and in this model, one can obtain a cosmological constant of just the right magnitude to give the observed expansion from the contribution of the ghost fields, which are supposed to be present in the low-energy effective theory of QCD without introducing any new degrees of freedom.  The ghosts are needed to solve the $U(1)$ problem, but they are completely decoupled from the physical sector \cite{Kawarabayashi:1980dp}. The ghosts make no contribution in the flat Minkowski space, but make a small energy density contribution to the vacuum energy due to the off-set of the cancellation of their contribution in curved space or time-dependent background. For example, in the Rindler space, the contribution of high frequency modes is suppressed by the factor $e^{-2\pi k/a_T}$ and the main contribution comes from $k\sim a_T$, where $a_T$ is the temperature on the horizon seen by the Rindler observer \cite{Ohta}. In the cosmological context,  one can choice $a_T\sim H$, which corresponds to the temperature on the Hubble horizon. Then, in the context of strongly interacting confining QCD with topological nontrivial sector, this effect occurs only in the time direction and their wave function in other space directions is expected to have the size of QCD energy scale. Thus, this ghost gives the vacuum energy density proportional to $\Lambda_{QCD}^{3} H_{0}$. With  $H_{0}\sim 10^{-33}$eV, it gives the right order of observed magnitude $\sim (3\times10^{-3}$eV)$^{4}$ of the energy density. 

It should be emphasized that the Veneziano ghost from the ghost dark energy model is not a new propagating physical degree of freedom and the description of dark energy in terms of the Veneziano ghost is just a matter of convenience to describe very complicated infrared dynamics of strongly coupled QCD and it does not violate unitarity, causality, gauge invariance and other important features of renormalizable quantum field theory, see \cite{Zhitnitsky:2010ji,Holdom:2010ak,Zhitnitsky:2011tr,Zhitnitsky:2011aa}.
 
 Generally, it is very difficult to accept the linear behavior that the energy of FRW universe is linear in Hubble constant "H", because QCD is a theory with a mass gap determined by the energy scale $100$MeV. So, it is generally expected that there should be an exponentially small corrections rather than that linear corrections H. However, as the arguments discussed in refs.\cite{Zhitnitsky:2010ji, Holdom:2010ak,Zhitnitsky:2011tr,Zhitnitsky:2011aa} that the linear scaling H is due to the complicated topological structure of strongly coupled QCD, not related to the physical massive propagating degree of freedom. Therefore, the linear in Hubble constant H scaling has a strong theoretical support tested in a number of models. For recent paper on fitting the ghost dark energy model with current observational data including SnIa, BAO, CMB, BBN, Hubble parameter and growth rate of matter perturbation, see \cite{Cai:2012fq} and also see \cite{RozasFernandez:2011je}. For recent progress on dark energy, see \cite{Feng:2009jr}.

Einstein equations can be written in a form called "unified first law" based on the general definition of black hole dynamics on trapping horizon, which was proposed by Hayward \cite{Hayward:1993wb}. There are some other horizons in the context of the FRW universe, such as the future inner trapping horizon defined in the next section, while the outer one is used to  define  black holes in a general spacetime including time-dependent spacetime \cite{Hayward:1993wb}. So, in this paper, we will choice $a_T\sim T_{t}$, where $T_{t}$ is the temperature defined on the trapping horizon of the  FRW universe instead to study its dynamical behavior locally and globally. In addition, we also study the bifurcation phenomenon in this dynamical system and predict the fate of the universe.

This paper is organized as follows: In Sec.\ref{sec::as}, we shall derive the equations of the autonomous system of the universe with  ghost dark energy component,  and investigate  its properties and solutions qualitatively and numerically in Sec.\ref{d2pp}.  The global behavior of this system will be presented in Sec.\ref{sec::gb}, and we also make a prediction of the fate of the universe in Sec.\ref{fou} and study its bifurcation phenomenon in Sec.\ref{sec::bir}.  In the last section, we will give some discussions and conclusions.

\section{Autonomous system}\label{sec::as}
The spacetime of our universe is described by the FRW metric, which could be written in the form of
\begin{equation}\label{FRW}
            ds^{2}=h_{ab}dx^{a}dx^{b}+\tilde{r}d\Omega^{2} \,,
\end{equation}
 where $x^{0}=t$, $x^{1}=r$ and  $\tilde{r}=a(t)r$, which is the radius of the sphere while $a(t)$ is the scale factor. Defining
 \begin{equation}
	   d\xi^{\pm}=-\frac{1}{\sqrt{2}}\left(dt \mp \frac{a}{\sqrt{1-kr^2}}dr\right) \,,
\end{equation}
where $k$ is the spacial curvature, the metric could be rewritten as a double-null form
\begin{equation}
	ds^2=-2 d\xi^+d\xi^- +\tilde{r}^2d\Omega^2 \,,
\end{equation}
then we get the trapping horizon $\tilde r_{T}$ by solving the equation $\partial_{+} \tilde r |_{\tilde r= \tilde r_{T}} = 0$ as
\begin{equation}\label{horizon_radus}
	\tilde r_{T} = \left(H^{2}+ \frac{k}{a^{2} } \right )^{-1/2} \,,
\end{equation}
which coincides with the apparent horizon \cite{Feng:2011ev}.  The surface gravity is given by
\begin{equation}\label{kappa}
	\kappa =-\frac{\tilde r_{T}}{2}\left(\dot H + 2H^{2}+\frac{k}{a^{2}}  \right)= -\frac{1}{\tilde r_{T}} \left(1-\epsilon\right) \,,
\end{equation}
where
\begin{equation}
	\epsilon \equiv \frac{\dot{\tilde r}_{T} }{2H\tilde r_{T}} = \frac{ (\ln \tilde r_{T})' }{2}\,,
\end{equation}
where dot and prime denote the derivative with respect to $t$ and $\ln a$, respectively. Here, we assume $\epsilon <1$ so that $\kappa<0$ corresponding to inner trapping horizon.

As we mentioned before, one can simply choice the temperature on the horizon as $T\sim H$ in the cosmological context, but this is a special choice and generally, in the context of dynamical spacetime, the temperature is well-defined on the trapping horizon, which is proportional to  the surface gravity on that horizon, namely  $T_t=|\kappa|/2\pi$. Actually, it will reduce to $T\sim H$ in the de Sitter spacetime when we neglect the spatial curvature, because the trapping horizon is coincident with the Hubble horizon in this situation. In the following, we will also take the idea of the ghost dark energy model, in which the vacuum energy density is proportional to the temperature defined on the trapping horizon $\rho_{DE} \sim T_t $. Therefore, the energy density of the Veneziano ghost dark energy is then given by
\begin{equation}
	\rho_{DE} = \frac{C}{\tilde r_{T}} \left(1-\epsilon\right) \,,
\end{equation}
where $C$ is a constant and here we have used Eq.~(\ref{kappa}). The Friedmann equation now reads
\begin{equation}\label{fried}
	H^{2} + \frac{k}{a^{2}}= \frac{1}{3} \big(\rho_{DE} + \rho_{m}\big) \,,
\end{equation}
where $\rho_{m}$ is the energy density of dark matter, and by solving the Friedmann equation, we get
\begin{equation}
	\rho_{DE} = \frac{C^{2}(1-\epsilon)^{2}}{6} \left[1+\sqrt{1+\frac{12\rho_{m}}{C^{2}(1-\epsilon)^{2}}}\right] \,.
\end{equation}

Usually, people also consider case that dark energy and dark matter are coupled to each other by adding a interaction term in the conservation equations~\cite{Wang:2005jx} . Of course, the entirely independent evolution of the dark energy is a special case with a vanished interaction term. Furthermore, the microphysics seems to allow enough room for the coupling between the two components. At leat, before drawing a definitely conclusion, the interaction case is still deserved to study. However, once  a generic interaction term  is introduced,  the dynamical equations would become much more complicated and can not be solved analytically. As we known, by using the locally dynamical analysis, one can get the future behavior of the nonlinear system and know whether the system is stable or not near the critical points. But, this method can only give the neighborhood properties of the critical points. So, in the following, we will develop a global method to study the global behavior of the nonlinear dynamical system, as well as its bifurcation phenomenon.
As we known, the conservation equations including a general interaction  between the dark energy and  dark matter is described by
\begin{eqnarray}
	\dot \rho_{m} + 3H \rho_{m} &=& Q \,, \label{e1} \\
	\dot \rho_{DE} + 3H(1+w_{DE})\rho_{DE} &=& -Q \label{e2}\,,
\end{eqnarray}
where $Q$ is the interacting term, and we will take its form as $Q=3H(\alpha \rho_{DE} + \beta \rho_{m})$ in the following. Here $\alpha$ and $\beta$ are  some constants. Usually, the existing literature only consider a simple case $\alpha = 0 $ or $\beta = 0$, as well as the case $\alpha=\beta$, see Refs.~\cite{Sheykhi:2011xz} and references therein.  We will see that, once we have developed a global analysis  method, it would be unnecessarily to set these special values, and there would be a interesting bifurcation phenomenon when one leaves these parameters free.

The equation of state parameter of dark energy then be derived as
\begin{equation}\label{eos}
	w_{DE} = \frac{1}{3}\left(\frac{\epsilon'}{1-\epsilon} + 2\epsilon -\frac{Q}{\rho_{DE} H}\right)-1 \,.
\end{equation}
By using Eqs.~(\ref{fried}), (\ref{e1}) and (\ref{e2}), we can obtain
\begin{equation}\label{f31}
	\dot H -\frac{k}{a^{2}}= -\frac{1}{2} \bigg( \rho_{DE}(1+w_{DE}) + \rho_{m}\bigg) \,,
\end{equation}
and we also have the relation
\begin{equation}\label{f32}
	\dot H -\frac{k}{a^{2}} = -\frac{2\epsilon}{\tilde r_{T}^{2}} \,.
\end{equation}

By using the following set of dimensionless variables
\begin{equation}\label{cons1}
	\Omega_{DE} = \frac{\rho_{DE}}{3H^{2}} \,,\quad \Omega_{m} = \frac{\rho_{m}}{3H^{2}} \,,\quad \Omega_{k} = \frac{-k}{a^{2}H^{2}} \,,
\end{equation}
then, the Friedmann equation (\ref{fried}) can be written as
\begin{equation}\label{constr}
	\Omega_{DE} + \Omega_{m}  + \Omega_{k} = 1 \,,
\end{equation}
which indicates that all terms in the above equation take values in the interval $[0,1]$.  The equation of state parameter  (\ref{eos}) could be rewritten in terms of new variables as
\begin{equation}
	w_{DE} = \frac{\epsilon'}{3(1-\epsilon)} + \frac{2\epsilon}{3} - \alpha - \beta \mu -1 \,,
\end{equation}
where $\mu$ is the ratio of energy density of dark energy to dark matter, i.e. $\mu=\Omega_{m}/\Omega_{DE}$,  and the deceleration parameter $q \equiv -\ddot a/(H^{2} a)$  is given by
\begin{equation}
	q = -1 +\Omega_{k} + 2\epsilon (1+\mu)\Omega_{DE} = -(1-2\epsilon)(1+\mu)\Omega_{DE} \,,
\end{equation}
where we have used the constraint (\ref{constr}) to get the last equality of the above equation. By using Eq.~(\ref{f31}) and (\ref{f32}), one can get the relation
\begin{equation}
	\frac{4\epsilon}{3} = 1+\frac{w_{DE} }{1+\mu}  \,.
\end{equation}
Thus, the dynamical equations can be written in the following form
\begin{eqnarray}
	\mu' &=& 3\bigg[\left(\beta-1+\frac{4\epsilon }{3}\right)\mu +\alpha \bigg](\mu+1)\,,  \label{d1}\\
	\epsilon' &=& (1-\epsilon) \bigg[2\epsilon + 3\alpha+3\left(\beta-1+\frac{4\epsilon}{3} \right)\mu\bigg] \,, \label{d2}
\end{eqnarray}
The Eqs.~(\ref{d1}) and (\ref{d2}) constitute an autonomous system and we shall study the properties of its solutions qualitatively and numerically, for some recent works on locally dynamic analysis see Ref.~\cite{Li:2009zzc}, and also see \cite{Li:2011sd} for a review.

Before ending this section, we would like to point out that the computation of the
speed of sound  should not be treated as a signal for
instability of the theory as the Veneziano ghost is not a propagating physical
degree of freedom. Such an "apparent signal for instability" is in fact a result
of treatment of the Veneziano ghost as the conventional physical degree of
freedom satisfying classical equation of motion. Appropriate interpretation
is that a notion of the speed of sound does not exist for a non-physical, non-
propagating degree of freedom. This viewpoint was adapted in recent paper, see \cite{RozasFernandez:2011je}.

\section{The picture of two-dimensional parameter plane}\label{d2pp}

The critical points of the autonomous system can be obtained by setting $\mu' = 0$ and $\epsilon' =0$. All the critical points are listed in coordinate $(\mu_{c}, \epsilon_{c})$ in Tab.\ref{table:cp}.
\begin{table}[h]
\centering
  \begin{tabular}{c|c|c|c}
  \hline
  \hline
  C.P.& $\mu_{c}$ & $\epsilon_{c}$ & P.C. ($\mu>0$)            	\\
  \hline
  $P_{1}$               &$ \frac{\alpha}{1-\beta}$                              & $0$ 		 & $\alpha >0, \beta <1$ or $\alpha<0, \beta >1$ \\
  \hline
  $P_{2}$               &$- \frac{\alpha}{\beta+\frac{1}{3}}$              & $1$ 		 & $\alpha<0, \beta>-1/3$ or $\alpha >0, \beta <-1/3$\\
  \hline
  $P_{3}$               &$-1$                              & $ \frac{3(\alpha-\beta +1) }{2}$  & always unphysical\\
  \hline
  $P_{4}$               &$-1$                              & $1$ 						& always unphysical\\
   \hline
  \hline
  \end{tabular}
  \caption{\label{table:cp} All the critical points (C.P.) and their physical conditions (P.C.). }
\end{table}
And one can see that if $\alpha >0, \beta <1$ or $\alpha<0, \beta >1$, then $P_{1}$ is accepted from physical condition, namely, $\mu \geq0$ by definition, and if $\alpha<0, \beta>-1/3$ or $\alpha >0, \beta <-1/3$,  then $P_{2}$ is accepted physically.  To investigate the property of these critical points and the bifurcation of two-parameter family $\alpha$ and $\beta$, we can write the variables near these critical points $(\mu_{c}, \epsilon_{c})$ in the form $\mu = \mu_{c} + \delta \mu$ and $\epsilon = \epsilon_{c} + \delta\epsilon$ with the perturbations $\delta \mu$ and $\delta \epsilon$, and thus we get the perturbation equations at the critical points $P_{i}$ as follows
\begin{equation}
\left(	\begin{array}{c}
 \delta\mu' \\
 \delta\epsilon'
\end{array}\right)\,
= A_{i}
\left(\begin{array}{c}
 \delta\mu \\
 \delta\epsilon
\end{array}\right)\,,
\end{equation}
where $A_{i}$ are the Jacobian matrices evaluated at the corresponding critical point $P_{i}$, see Appendix \ref{jm} for explicit forms of these matrices.

\begin{figure}[!htbp]
\centering
\includegraphics[height=0.4\textwidth,width=0.4\textwidth]{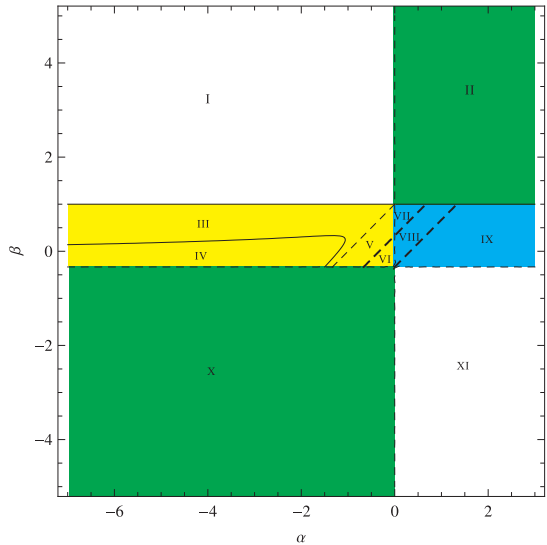}
\caption{The two-parameter plane is divided into 11 domains in accordance with the type of critical point and the physical condition $\mu \leq 0$. All the critical points have same property in each domain.}
 \label{Centera1g0}
\end{figure}

The character equation of $A_{i}$ can be written as
\begin{equation}
	\lambda^{2} - (\text{tr} A_{i}) \lambda + \det A_{i} = 0 \,,
\end{equation}
and the location of $(\text{tr}A_{i}, \det A_{i})$ relative to the parabola $(\text{tr}A_{i})^{2}-4\det A_{i} = 0$ in the trace-determinant plane determines the property of critical points $P_{i}$.  A $2\times 2$ matrix can be regarded as a $4$-dimensional space  determined by $4$ elements of the matrix. The trace-determinant plane is a $2$-dimensional representation of the $4$-dimensional space. Thus, there are infinite different matrices corresponds to each point in the trace-determinant plane. Furthermore,  each point in the $\alpha$-$\beta$ plane corresponds to some point in the trace-determinant plane, since $A_{i}$ depends on the parameters $\alpha$ and $\beta$. In other words, the property of critical points $P_{i}$ may change with the parameters $\alpha$ and $\beta$, because the family of dynamical system (\ref{d1}) and (\ref{d2}) depends on $\alpha$ and $\beta$.  By using a lengthy and straight computing, we can divide the $\alpha$-$\beta$ plane into $11$ domains (see Fig.\ref{Centera1g0}) where all the critical points have same type in each domain. In Tab. \ref{table:cpp}, all the  critical points and their properties are listed. Noteworthily, the property of critical point is obtained from the linearized system of the nonlinear system near the critical point in the Tab. \ref{table:cpp}. We say that a critical point $P_{i}$ of a nonlinear system is hyperbolic if the eigenvalues of linearized system have nonzero real parts. The linearization theorem show that the nonlinear flow is conjugate to the flow of the linearized system in a neighborhood of $P_{i}$. Therefore, there properties are preserved for the nonlinear system (\ref{d1}) and (\ref{d2}). In the domains I and XI, there are two physical critical points $P_{1}$ and $P_{2}$, and there is a physical point $P_{2}(P_{1})$ in the domain III-VI (VI-IX). II and X are domains without physical points.

\begin{table}[h]
\centering
  \begin{tabular}{c|c|c|c|c}
  \hline
  \hline
  Domain    & $P_{1}$            & $P_{2}$            & $P_{3}$            & $P_{4}$            	 \\
  \hline
  \hline
  I               & source(p.c.p)    & source(p.c.p)    & saddle              & saddle    		 \\
  \hline
  II              & \multicolumn{4}{c}{no physical critical point}   \\
   \hline
  III             & sink                   & saddle(p.c.p)    & saddle              & saddle    		 \\
   \hline
  IV             & spiral sink         & saddle(p.c.p)    & saddle              & saddle   		 \\
   \hline
  V              & saddle   	     & saddle(p.c.p)    & sink                   & saddle   		 \\
   \hline
  VI             & saddle              & saddle(p.c.p)    & saddle              & sink   			 \\
   \hline
  VII            & saddle(p.c.p)    & saddle              & sink                  & saddle   		 \\
   \hline
  VIII           & saddle(p.c.p)    & saddle              & saddle              & sink   			 \\
   \hline
  IX             &saddle(p.c.p)     & sink                  & saddle              & saddle   		 \\
   \hline
  X              & \multicolumn{4}{c}{no physical critical point}    \\
   \hline
  XI             & saddle(p.c.p)    & sink(p.c.p)        & saddle              & saddle    \\
   \hline
  \hline
  \end{tabular}
  \caption{\label{table:cpp} The critical points of the autonomous system (\ref{d1}) and (\ref{d2}) and their type in each domain of the $\alpha$-$\beta$ plane. }
\end{table}

In general, a bifurcation occurs when there is a remarkably change in the structure of the solutions of the dynamical system as the parameters $\alpha$ and $\beta$. The simplest types of bifurcation occurs when the property of equilibrium solutions changes as $\alpha$ and $\beta$ vary. Thus, a bifurcation may be appearance when the parameters vary from some domain into neighbor one. For example, the structure of $P_{1}$-$P_{2}$ changes from saddle-saddle to saddle-sink and the structure of $P_{3}$-$P_{4}$ changes from sink-saddle to saddle-saddle when the location of $(\alpha,\beta)$ crosses the line $3\alpha-3\beta-1=0$, namely, $(\alpha,\beta)$ moves from domain VIII to IX in the $\alpha$-$\beta$ plane.

\section{Global behavior of the nonlinear system}\label{sec::gb}

In the section \ref{d2pp}, we have used the technique of linearization to determine the behavior of solutions near physical points. In this section, we use the qualitative technique of nullcline for analyzing the global behavior of nonlinear system, which is one of the most useful tools for analyzing solutions of nonlinear planar system. For our system, the $\mu$-nullclines are the set of points determined by
\begin{equation}\label{munull}
	\mu = -1; \quad \left(\beta-1+\frac{4\epsilon}{3}\right) \mu = -\alpha \,.
\end{equation}
Similarly, the $\epsilon$-nullclines are the set of points determined by
\begin{equation}\label{epnull}
	\epsilon = 1; \quad \left(\beta-1+\frac{4\epsilon}{3}\right) \left(\mu +\frac{1}{2}\right)= \frac{1}{2}(\beta-1)-\alpha \,.
\end{equation}
It is worth noting that the nullclines consist of  lines and hyperbolas. The asymptotes of hyperbola in Eq.~(\ref{munull}) are $\mu=0$ and $\epsilon=1-\beta$, which is interesting to determine the fate of universe.

On the $\mu$-nullclines, we have $\mu'=0$, the vector field points $(\mu', \epsilon')$ are vertical, so $\mu$-nullclines divide $\mu$-$\epsilon$ plane into regions where the vector field points either to the left or to the right. Similarly, the $\epsilon$-nullclines separate the plane into regions where the vector field points either upward or downward. If we determine all of the nullclines, then this allows us to decompose $\mu$-$\epsilon$ plane into a collection of basic regions. The vector field must point in one of $4$ directions (northeast, northwest, southeast or southwest) for any of the basic regions between the nullclines, since it is neither vertical  nor horizontal. In each of basic regions, the vector field points are provided with a certain direction. Therefore, the basic regions allow us to understand the phase portrait from a qualitative point of view. Furthermore, the nullclines are depending on the parameters $\alpha$ and $\beta$ in our dynamical system, so we have to discuss each of domains I-XI separately.

\begin{figure}[!htbp]
\centering
\includegraphics[height=0.4\textwidth,width=0.4\textwidth]{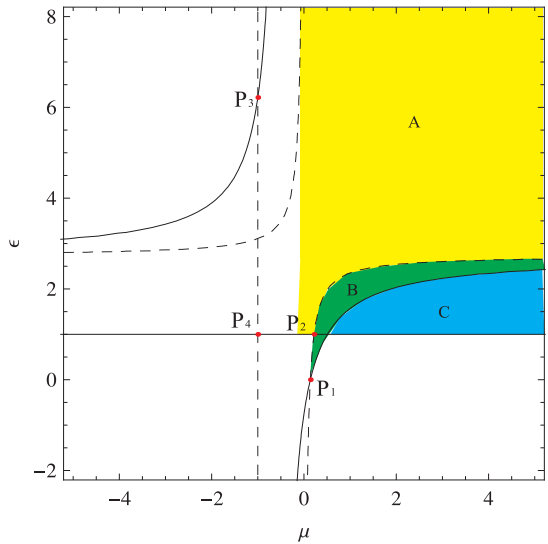}
\caption{The $\mu$-$\epsilon$ plane are divided into 13 basic regions by the nullclines as $\alpha, \beta \in$ Domain I. The direction of the vector field at all points in the basic region are determined.} \label{fig2a}
\end{figure}

\begin{figure}[!htbp]
\centering
\includegraphics[height=0.4\textwidth,width=0.4\textwidth]{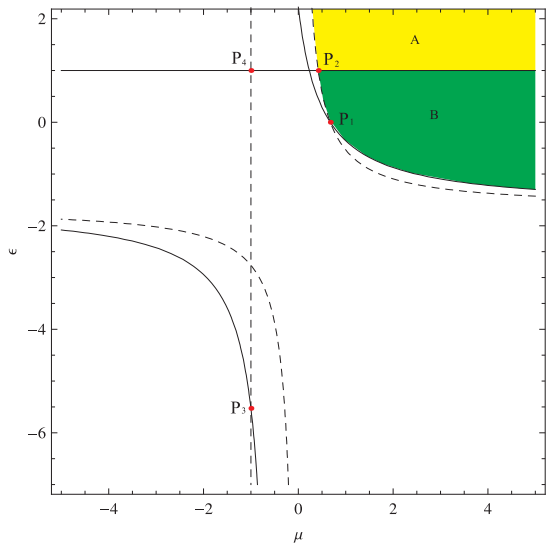}
\caption{The $\mu$-$\epsilon$ plane are divided into 13 basic regions by the nullclines as $\alpha, \beta \in$ Domain XI. We can decide the direction of the vector field at all points in the basic region.}\label{fig2b}
\end{figure}

Here, we discuss domain I and XI in detail, since there are two physical critical points $P_{1}$ and $P_{2}$ in these domains. In the case of $\alpha, \beta \in$ Domain I, the nullclines divide $\mu$-$\epsilon$ plane into $13$ basic regions in Fig.\ref{fig2a}. By taking one point in each of these regions firstly and then deciding the direction of the vector field at that point, we can determine the direction of the vector field at all points in the basic region. Similarly, the nullclines also divide  $\mu$-$\epsilon$ plane into $13$ basic regions for the case of $\alpha, \beta \in$ Domain XI, see Fig.\ref{fig2b}, and we can decide the direction of the vector field at all points in the basic region. Especially, we can find out the approximate behavior of solutions everywhere in the plane, which has interest for the discussion of physical significance. For example, we will know that the matter component is a dominant position at late evolution if any solution tends toward infinity in the basic region marked A(B) in Fig.\ref{fig2b} with the northeast (southeast) direction. In other words, we will know the fate of universe and its initial conditions.


By using above analysis and the numerical calculation, we can show the global behavior of the nonlinear system exactly. In Fig.\ref{fig3}, we plot evolution trajectories in the $\mu$-$\epsilon$ plane where we have chosen the parameters $\alpha, \beta\in$ Domain XI. We also plot evolution trajectories when $\alpha, \beta\in$ Domain I in Fig.\ref{fig4}. It is worth noting that the physical admissible range is the only right-half plane ($\mu\geq0$), so we must choose initial conditions at the right-half plane. The detailed discussion of the physical arguments will appear in Section \ref{fou}.

\section{Fate of universe}\label{fou}
\begin{figure}[!htbp]
\centering
\includegraphics[height=0.4\textwidth,width=0.4\textwidth]{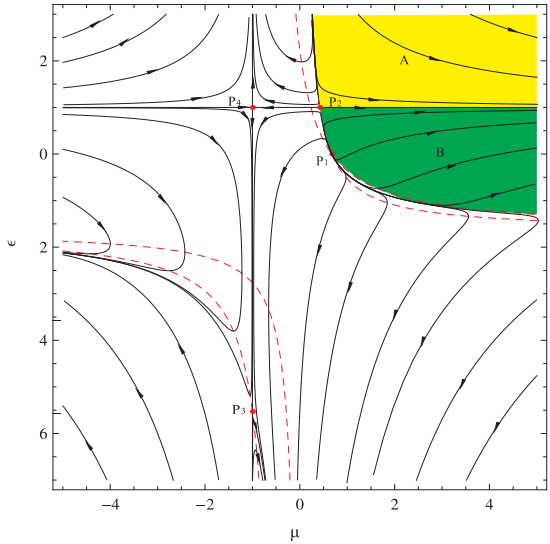}
\caption{Choosing $\alpha, \beta$ in domain XI, the evolution trajectories in the $\mu$-$\epsilon$ plane are plotted.} \label{fig3}
\end{figure}

From Fig.\ref{fig3}, one can see that if $\alpha, \beta$ belongs domain I, and the initial values of $\mu$ and $\epsilon$ satisfy the following condition
\begin{eqnarray}
	\left(\beta -1 + \frac{4\epsilon_{0}}{3}\right)\mu_{0} +\alpha &>& 0 \,, \\
	3\left(\beta -1 + \frac{4\epsilon_{0}}{3}\right)\mu_{0} + 2\epsilon_{0} + 3\alpha &>& 0   \,,
\end{eqnarray}
i.e. the initial point ($\mu_{0}, \epsilon_{0}$) belongs to A $\bigcup$ B (A denotes the yellow region, while B denotes the green region), the universe will evolve to matter dominate finally, which suggests that the expansion will asymptotically come to a halt. If $\alpha, \beta$ belongs domain I, and the initial value of $\mu$ and $\epsilon$ satisfy the following condition
\begin{eqnarray}
	\left(\beta -1 + \frac{4\epsilon_{0}}{3}\right)\mu_{0} +\alpha &<& 0 \,, \\
	3\left(\beta -1 + \frac{4\epsilon_{0}}{3}\right)\mu_{0} + 2\epsilon_{0} + 3\alpha &<& 0   \,,
\end{eqnarray}
the dynamical evolution will be in unphysical region, so such kind of initial values are excluded by the physical condition $\mu \geq0$.

\begin{figure}[!htbp]
\centering
\includegraphics[height=0.4\textwidth,width=0.4\textwidth]{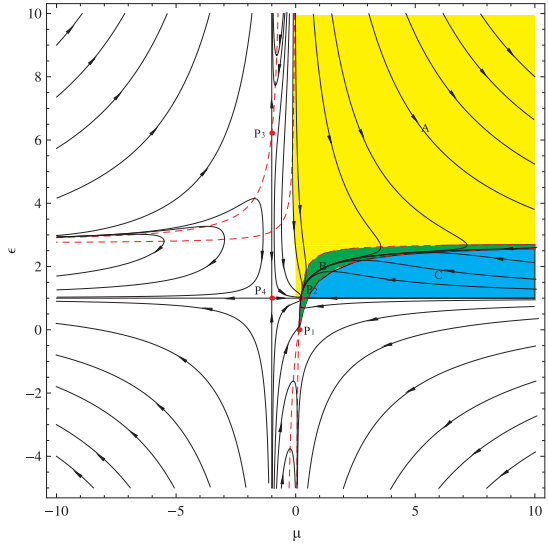}
\caption{We plot the evolution trajectories in the $\mu$-$\epsilon$ plane as $\alpha, \beta$ are in domain XI.} \label{fig4}
\end{figure}

From Fig.\ref{fig4}, one can see that if $\alpha, \beta$ belongs domain XI, and if not only the initial value of satisfies the physical condition $\mu \geq 0$, but also satisfies $\epsilon_{0} \geq 1$, then,  the system will evolute to $P_{2}$ finally, namely, the fate of universe will be
\begin{equation}
	\frac{\Omega_{m}}{\Omega_{DE}} = -\frac{\alpha}{\beta + \frac{1}{3}} \,, \quad (\alpha>0\,, \beta <-\frac{1}{3}) \,.
\end{equation}
If we assumed that we have get the above limit at present, then, we can choose $\alpha$ and $\beta$ as
\begin{equation}
	\beta \approx -\frac{1}{3} - \alpha \frac{\Omega_{DE,0}}{\Omega_{m,0}} \,,
\end{equation}
where $\Omega_{DE,0}$ and $\Omega_{m,0}$ denote the present values of energy density  of dark energy and matter respectively.  If $\alpha, \beta$ belongs to domain XI, and the initial values satisfy $\epsilon \leq 0$ with physical condition $\mu \geq 0$, the dynamical system will tend to be $\mu\rightarrow 0$, namely, the universe will be dominated by dark energy, and will be accelerating eternally.


\section{Bifurcation}\label{sec::bir}
The bifurcation phenomenon occurs when there is a remarkable change in the structure of the solutions of the dynamical system as the parameter $\alpha$ and $\beta$. Here, we consider that the structure of the solutions makes a change as the parameter $\beta$ crosses line $\beta =1$.

\begin{figure}[!htbp]
\centering
\includegraphics[height=0.4\textwidth,width=0.4\textwidth]{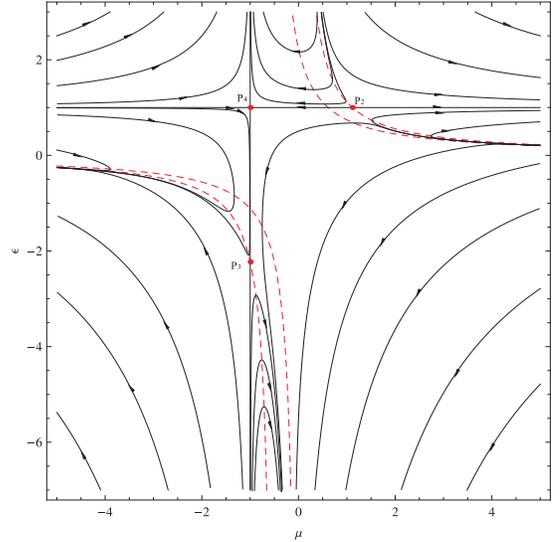}
\caption{As $\alpha = -1.485$ and $\beta =1$, we plot the evolution trajectories in the $\mu$-$\epsilon$ plane.}\label{fig5}
\end{figure}

\begin{figure}[!htbp]
\centering
\includegraphics[height=0.4\textwidth,width=0.4\textwidth]{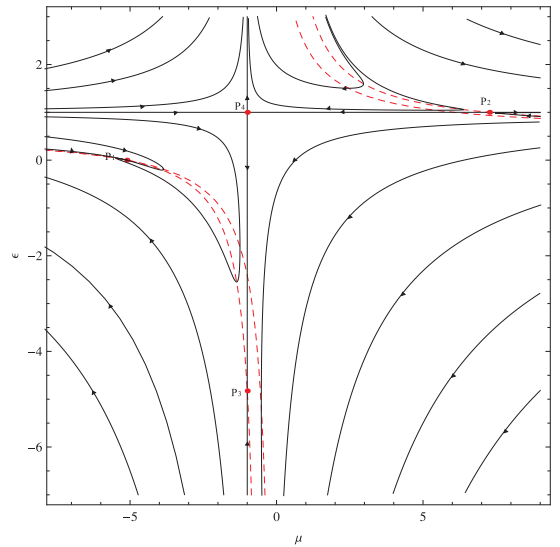}
\caption{Fixing $\alpha = -4$ and $\beta = 0.217$, we plot the evolution trajectories in the $\mu$-$\epsilon$ plane.}\label{fig6}
\end{figure}

Now, we consider the case $\alpha<0$ and $\beta=1$. The system is simplified to
\begin{eqnarray}
	\mu'& =& (4\epsilon \mu + 3\alpha)(\mu +1) \,, \\
	\epsilon' &=& (2\epsilon + 3\alpha + 4\epsilon \mu)(1-\epsilon) \,.
\end{eqnarray}
In this case, there are only three critical points: $P_{2} ( -3\alpha/4,1)$, $P_{3}(-1,3\alpha/2)$ and  $P_{4}(-1,1)$ are all saddle points, and the physical discussion is not change except losing physical point $P_{1}$. In Fig.\ref{fig5}, we plot evolution trajectories in the $\mu$-$\epsilon$ plane for $\alpha = -1.485$ and $\beta =1$. For the parameters $\alpha, \beta\in$ Domain III, we also plot evolution trajectories in   the $\mu$-$\epsilon$ plane for $\alpha = -4$ and $\beta =0.217$ in Fig.\ref{fig6}. Obviously, $P_{1}$ changes into a sink from a source as the parameters $\alpha, \beta$ run in Domain II from Domain I, and $P_{1}$ also becomes a non-physical point. Therefore, one can say that this is a typical bifurcation for the nonlinear system. However, the physical arguments of evolution have not remarkable change.


\section{Conclusion}
In conclusion, we have studied the dynamical behavior of the universe accelerated by the generalized  Veneziano ghost dark energy  component locally and globally. We have found that in this system, there are four critical point but only two of them have physical meaning when the parameters are chosen in some proper regions to satisfy the physical conditions.  We have used the technique of linearization to determine the behavior of solutions near physical points and used the qualitative technique of nullcline for analyzing the global behavior of this nonlinear system, which is one of the most useful tools for analyzing solutions of nonlinear planar system. We have shown that the universe could be dominated by the dark energy at present if we choice a set of suitable parameters. We also give an example to show the bifurcation phenomenon in this interesting dynamical system. Before ending the paper, we would like to emphasis that the  qualitative technique of nullcline we developed in this paper is very powerful and  could be used in any nonlinear dynamical system, especially in the planar system, so it deserves further studying.

\acknowledgments

This work is supported by National Science Foundation of China grant Nos.~11105091 and~11047138, National Education Foundation of China grant  No.~2009312711004, Shanghai Natural Science Foundation, China grant No.~10ZR1422000, Key Project of Chinese Ministry of Education grant, No.~211059,  and  Shanghai Special Education Foundation, No.~ssd10004.

\appendix

\section{Jacobian matrixes} \label{jm}

\begin{equation}       
A_{1}=
\left(
\begin{array}{ccc}
 -3 \left(1+\alpha-\beta \right) && \frac{4 \alpha \left(1+\alpha-\beta\right)}{\left(1-\beta\right)^2} \\
 \\
 3 \left(\beta-1\right) && \frac{2 \left(\beta-2\alpha-1\right)}{\beta-1}
\end{array}
\right)\,.
\end{equation}

\begin{equation}       
A_{2}=
\left(
\begin{array}{ccc}
1-3\alpha + 3\beta && \frac{12\alpha(3\alpha-3\beta-1)}{\left(1+3\beta\right)^2} \\
 \\
0 && -2
\end{array}
\right)\,.
\end{equation}

\begin{equation}       
A_{3}=
\left(
\begin{array}{ccc}
-3(1+\alpha -\beta) && 0\\
 \\
-\frac{3}{2}(1+3\alpha-3\beta)(1+2\alpha-\beta) && 1+3\alpha-3\beta
\end{array}
\right)\,.
\end{equation}

\begin{equation}       
A_{4}=
\left(
\begin{array}{ccc}
-1+3\alpha - 3\beta && 0 \\
 \\
0 && -1-3\alpha+3\beta
\end{array}
\right)\,.
\end{equation}

\end{document}